\documentclass[11pt,twoside]{article}
\usepackage{./asp2023}
\usepackage{placeins}
\usepackage{graphicx}

\aspvoltitle{Compendium of Undergraduate Research in Astronomy and Space Science}       

\resetcounters
\begin{document}
\title{Exploring Barred Galaxies in the Young Universe at $z \sim2$ Using \textit{JWST} CEERS Data}
\author{Keith Pritchett Jr.,$^1$ Shardha Jogee,$^2$ and Yuchen Guo$^3$
\vspace{1mm}
\affil{$^1$Siena College, Loudonville, NY, USA; \email{ka28prit@siena.edu}}
\affil{$^2$The University of Texas, Austin, TX, USA; \email{sj@astro.as.utexas.edu}}
\affil{$^3$The University of Texas, Austin, TX, USA; \email{kayguo98@utexas.edu}}}

\paperauthor{Sample$\sim$Author1}{Author1Email@email.edu}{ORCID_Or_Blank}{Author1 Institution}{Author1 Department}{City}{State/Province}{Postal Code}{Country}
\paperauthor{Sample$\sim$Author2}{Author2Email@email.edu}{ORCID_Or_Blank}{Author2 Institution}{Author2 Department}{City}{State/Province}{Postal Code}{Country}
\paperauthor{Sample$\sim$Author3}{Author3Email@email.edu}{ORCID_Or_Blank}{Author3 Institution}{Author3 Department}{City}{State/Province}{Postal Code}{Country}

\begin{abstract} \label{abstract}
Studying barred galaxies at early epochs can shed light on the early evolution of stellar bars, their impact on secular evolution and the star formation activity of young galaxies, and the origins of present-day barred galaxies like the Milky Way. We analyze data from the James Webb Space Telescope (\textit{JWST}) Cosmic Evolution Early Release Science (CEERS) Survey to explore the impact of rest-frame wavelength and spatial resolution on detecting and characterizing some of the highest redshift barred galaxies known to date, at $z\sim2$ corresponding to an epoch when the universe was only $\sim22\%$ of its current age. We apply both visual classification and ellipse-fitting to \textit{JWST} images (including F115W, F200W, and F444W images) of the barred galaxy CEERS-30155 at $z\sim2.136$. We find that the stellar bar in CEERS-30155 is not visible in the F115W image, which traces rest-frame ultraviolet (UV) light at $z\sim2$, a rest-frame wavelength highly obscured by dust. The stellar bar is visible in the F200W image, but is most prominent in the F444W image, likely due to the F444W image tracing rest-frame near-infrared (NIR) light at $z\sim2$. Rest-frame NIR light is not obscured by dust and traces low-mass, long-lived stars that dominate the stellar mass in galaxies. However, ellipse fits of the F444W image only robustly detect stellar bars whose semimajor axis (sma) are at least one PSF ($\sim0\farcs16$ or $\sim1.4$ kpc at $z\sim2$). At $z\sim2$, stellar bars smaller than 1.5 kpc will be more robustly detected in the sharper F200W image (PSF $\sim0\farcs08$ or $\sim0.7$ kpc at $z \sim2$), provided that the rest-frame optical light it traces is not overly impacted by dust and can still unveil the bar structure. Using a combination of both \textit{JWST} F200W and F444W images can improve the detection of barred galaxies at $z\sim2$ to 4. At even higher redshifts ($z>4$), the Giant Magellan Telescope with adaptive optics will be a cornerstone facility to explore young barred galaxies.
\end{abstract}

\section{Introduction} \label{intro}
\par
Stellar bars play a crucial role in the secular evolution of disk galaxies by driving gas from the outer disk to the circumnuclear region, enhancing central star formation and building disky bulges. In the present-day universe, $\sim60\%$ of massive disk galaxies are barred (e.g., \citealt{Athanassoula2003, Kormendy&Kennicutt2004, Jogee-Scoville-Kenney2005, Sellwood2016}), including the Milky Way (\citealt{Peters1975, Blitz-Spergel1991, Weiland-etal-1994}). Studying stellar bars at earlier epochs can provide valuable insights into how they may have influenced secular evolution and enhanced the central star formation rate of disk galaxies. Previous stellar bar studies that used data from the Hubble Space Telescope (\textit{HST}) primarily explored barred galaxies at $z<1.5$ (e.g., \citealt{Jogee-etal-2004, Melvin-etal-2014, Simmons-etal-2014}). These studies could not robustly detect stellar bars in galaxies at $z\gtrsim2$ because stellar bar structures at these epochs can be obscured by dust and/or be too small to resolve with the spatial resolution and wavelength of \textit{HST} ACS or WFC3 images. With the advent of high-resolution \textit{JWST} NIRCam images, the study of stellar bars has been extended to $z\gtrsim2$ (e.g., \citealt{Guo-etal-2023, Guo-etal-2024, Le-Conte-etal-2024}). New observational results on barred galaxies at $z\gtrsim2$ from \textit{JWST} data are now sharply constraining predictions by cosmological simulations (e.g., \citealt{Rosas-Guevara-etal-2020, Rosas-Guevara-etal-2022, Bi-Shlosman-Romano-Diaz2022}) and sparking interesting discussions about early stellar bar formation and disk galaxy evolution (e.g., \citealt{Bland-Hawthorn-etal-2023}).

\par
The primary goal of our study is to explore the factors that influence stellar bar detection in high-redshift disk galaxies, focusing on the impact of rest-frame wavelength and spatial resolution. We investigate how these factors impact the identification of barred galaxies by analyzing CEERS-30155, a barred galaxy at $z\sim2.136$ from \cite{Guo-etal-2023}, through a combination of visual classification and quantitative analysis via ellipse fits of  different \textit{JWST} images. CEERS-30155 is one of the highest redshift barred galaxies known to date, at an epoch when the universe was $\sim22\%$ of its current age. Our work provides insight into the methods of detecting high-redshift barred galaxies, and helps pave the way for larger studies that can assess the early evolution of stellar bars, their impact on secular evolution and the star formation activity of young galaxies, and the origins of present-day barred galaxies like the Milky Way.

\section{Methodology} \label{methods} 
We analyzed the NIRCam images of CEERS-30155 from the \textit{JWST} CEERS Survey (\citealt{Finkelstein-etal-2022, Bagley-etal-2023}). CEERS-30155 has a spectroscopic redshift of $z\sim2.136$, corresponding to $\sim22\%$ of the age of the universe (\citealt{Brammer-etal-2012}). The selected galaxy has been classified as barred using the F444W image in \cite{Guo-etal-2023}.

\par 
We use visual classification and ellipse fitting to identify the stellar bar structure in the \textit{JWST} F115W, F150W, F200W, F277W, F356W, and F444W images, respectively. We also use these methods to identify the stellar bar in the \textit{HST} F160W image. For CEERS-30155, the \textit{JWST} images trace rest-frame wavelengths ranging from UV ($\sim3700$ \AA) in F115W images to optical ($\sim6400$ \AA) in F200W images, and NIR ($\sim1.42$ $\mu m$) in F444W images. The \textit{JWST} images have a PSF FWHM ranging from $\sim0\farcs07$ in F115W images to $\sim0\farcs16$ in F444W images, corresponding to a range from 0.6 kpc to 1.4 kpc (Table \ref{tab:tel_bands} and Figure \ref{fig:CEERS_30155_Panel}). 
\par
Visually, we identify stellar bars as elongated stellar features that extend from the center of a galaxy to the outer disk and are often connected to spiral arms. We perform ellipse fitting on each image, generating radial profiles of surface brightness, ellipticity (\textit{e}), and position angle (PA) (Figure \ref{fig:ellipse_fit}). For stellar bar identification, we overlay ellipses on the image to see if the fitted ellipses are driven by the stellar bar feature in the galaxy. We consider a stellar bar to be present if the following criteria based on orbital structure are met: (i) Along the stellar bar, ellipticity (\textit{e}) steadily increases to a maximum value greater than 0.25 while the position angle remains nearly constant; (ii) At the transition from the bar structure to the outer disk, \textit{e} must drop by at least 0.1 (e.g., \citealt{Jogee-etal-2004, Marinova&Jogee2007, Guo-etal-2023, Guo-etal-2024}).

\FloatBarrier
\begin{table}[!h]
    \smallskip
    \centering
    {\small
    \begin{tabular}{cccccc}
    \tableline
    \noalign{\smallskip}

         Telescope & IR Image & $\lambda_{obs}$ ($\mu m$) & PSF ('') & $\lambda_{rest}$ at $z\sim2$ ($\AA$) & PSF at $z\sim2$ (kpc) \\

    \noalign{\smallskip}
    \tableline
    \noalign{\smallskip}
    
         Hubble & F160W & 1.60 & \multicolumn{1}{c|}{0.18} & 5100 & 1.53 \\
         Webb & F115W & 1.15 & \multicolumn{1}{c|}{0.07} & 3700 & 0.58 \\
         Webb & F150W & 1.50 & \multicolumn{1}{c|}{0.07} & 4800 & 0.61 \\
         Webb & F200W & 2.00 & \multicolumn{1}{c|}{0.08} & 6400 & 0.69 \\
         Webb & F277W & 2.77 & \multicolumn{1}{c|}{0.13} & 8800 & 1.09 \\
         Webb & F356W & 3.56 & \multicolumn{1}{c|}{0.14} & 11400 & 1.22 \\
         Webb & F444W & 4.44 & \multicolumn{1}{c|}{0.16} & 14200 & 1.39 \\

    \noalign{\smallskip}
    \tableline
    \end{tabular}
    
    \caption{The table shows the observed wavelength and PSF of different \textit{JWST} and \textit{HST} images used in this study, as well as corresponding rest-frame wavelengths and spatial resolutions at $z\sim2$. The \textit{HST} F160W image has the worst PSF. The \textit{JWST} images have better PSFs and as we go from F115W to F200W to F444W, they trace rest-frame wavelengths that increase from rest-frame UV to rest-frame optical to rest-frame NIR at $z\sim2$, respectively.}
    \label{tab:tel_bands}
    }
\end{table}
\FloatBarrier

\section{Results \& Discussion} \label{results}
\par
The detection and characterization of barred galaxies in the early universe is dependent on many factors, including rest-frame wavelength, spatial resolution, image depth, and methods to identify stellar bars. Longer wavelength NIR light can trace low-mass, long-lived stars that dominate the stellar mass of the stellar bar structure. A sharp PSF is essential for ellipse fits to show the stellar bar detection signature (see Section \ref{methods}), and stellar bars with a semi-major axis less than the image PSF are not robustly detected via ellipse fits. 

\FloatBarrier
\begin{figure}[h]
    \centering
    \includegraphics[width = \linewidth]{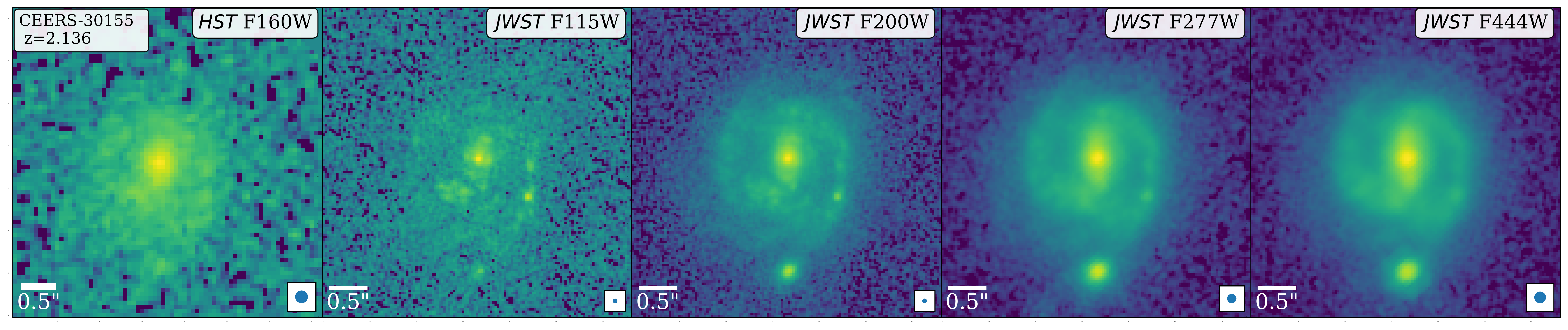}
    \caption{Montage of \textit{HST} and \textit{JWST} images for the galaxy CEERS-30155 at a redshift of $z \sim2.136$. From left to right, the \textit{JWST} F115W, F200W, F277W and F444W images have increasingly longer observed infrared wavelengths (1.15 to 4.44 $\mu m$) and trace light at rest-frame wavelengths ranging from UV to optical to NIR (Table \ref{tab:tel_bands}). The bar is strikingly revealed in F444W image, which traces rest-frame NIR light that probes the stellar mass of the galaxy and is less impacted by dust. The PSF of the \textit{JWST} images varies from $0\farcs07$ to $0\farcs16$, corresponding to a range from 0.6 kpc to 1.4 kpc at $z\sim2$. The blue circle at the bottom right of each image represents the PSF and the horizontal bar at the bottom left of each image shows a $0\farcs5$ scale for referencing.}
    \label{fig:CEERS_30155_Panel}
\end{figure}
\FloatBarrier

\par
We discuss the impact of rest-frame wavelength and spatial resolution in F115W, F200W, and F444W images:

\FloatBarrier
\begin{figure}[h]
    \centering
    \includegraphics[width=\linewidth]{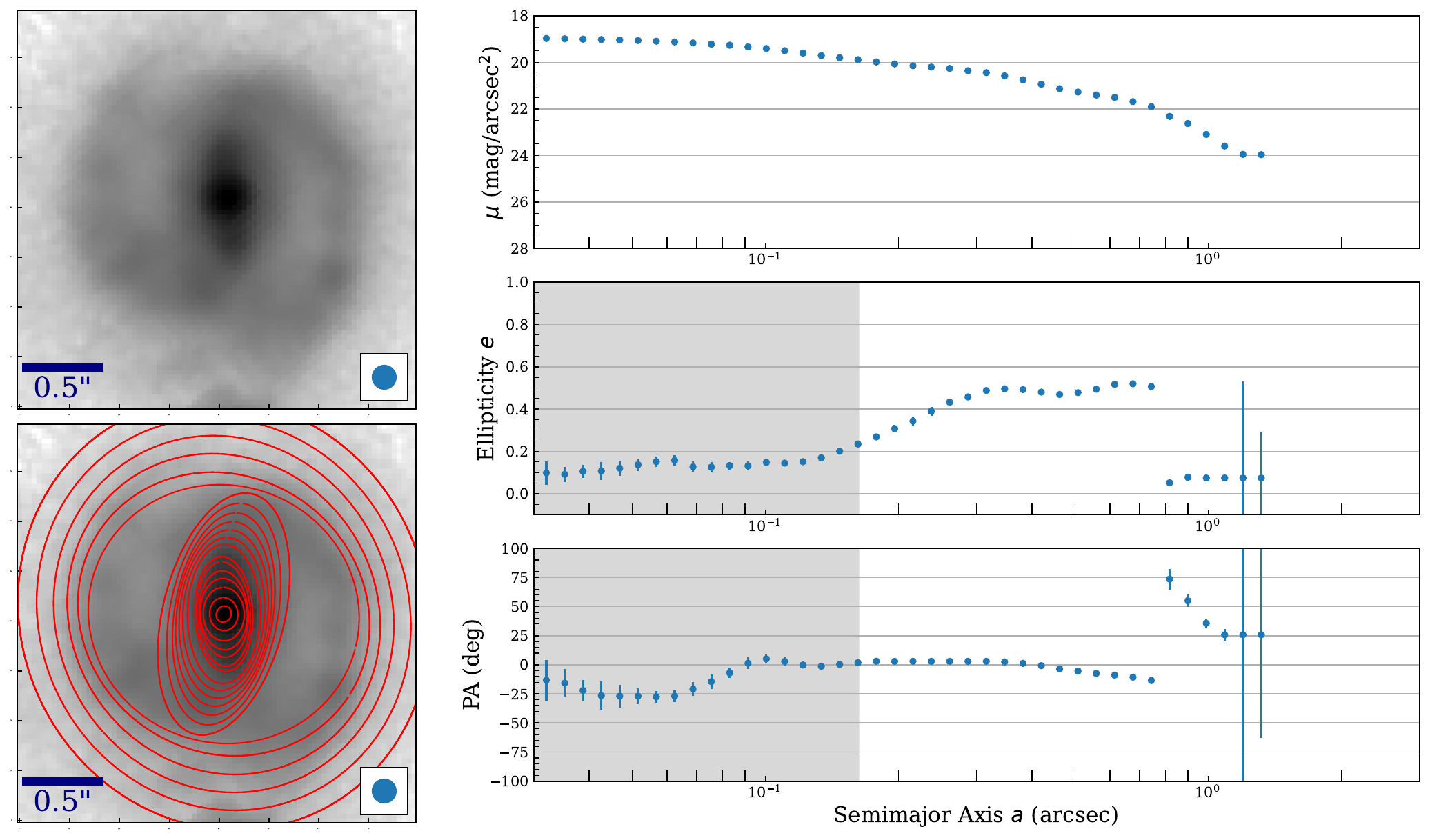}
    \caption{Ellipse fits of the \textit{JWST} F444W image of CEERS-30155 at $z\sim2.136$. The image traces rest-frame NIR light. The left images show the galaxy (top) and ellipse fits overlaid upon the image (bottom), with the blue circle at the bottom right of each image representing the PSF and the horizontal bar at the bottom left of each image showing a $0\farcs5$ scale for referencing. The right panels show the radial profiles of surface brightness, ellipticity (\textit{e}), and position angle (PA), with the light gray region representing one PSF. These profiles meet the quantitative criteria for stellar bar identification via ellipse fits described in Section \ref{methods}. The stellar bar can be identified both visually and via ellipse fits.}
    \label{fig:ellipse_fit}
\end{figure}
\FloatBarrier

\begin{itemize}
\item \textbf{F115W Image}: The stellar bar is not at all evident in visual classification (Figure \ref{fig:CEERS_30155_Panel}) or ellipse fits of the F115W image. At $z\sim2$, the short-wavelength F115W image does not reveal the stellar bar as it traces rest-frame UV light, which probes high-mass, short-lived stars and is obscured by dust.

\item \textbf{F200W Image}: The stellar bar of size $\sim2.9$ kpc is visible in the F200W image, but it is much more prominent in the visual classification and ellipse fits of the F444W image (Figures \ref{fig:CEERS_30155_Panel} \& \ref{fig:ellipse_fit}). The F200W image is able to reveal this stellar bar at $z\sim2$ due to two factors: its PSF of $\sim0\farcs08$ ($\sim0.7$ kpc) can resolve the stellar bar, and the rest-frame optical light it traces is not overly impacted by dust and can still unveil the stellar bar structure. In general, the F200W image is sharper (PSF $\sim0\farcs08$) than the F444W image (PSF $\sim0\farcs16$) and is well suited for resolving small structures. At $z\sim2$, stellar bars smaller than 1.5 kpc will not be robustly detected in F444W images, but can be revealed by F200W images provided that they are not extremely obscured by dust.

\item \textbf{F444W Image}: The stellar bar in CEERS-30155 is most prominent in the visual classification and ellipse fits of the F444W image (Figures \ref{fig:CEERS_30155_Panel} \& \ref{fig:ellipse_fit}) than of the F115W and F200W images. Ellipse fits of the F444W image shows that the stellar bar has an ellipticity (\textit{e}) of 0.5, and a moderately large semi-major axis (sma) of $\sim0\farcs35$, corresponding to $\sim2.9$ kpc at $z\sim2$ (Figure \ref{fig:ellipse_fit}). At $z\sim2$, the F444W image effectively reveals this moderately size stellar bar because of two factors: its PSF ($\sim0\farcs16$ corresponding to 1.4 kpc at $z\sim2$) can resolve the bar, and it traces rest-frame NIR light, which is not obscured by dust and traces low-mass, long-lived stars that dominate the stellar mass of the galaxy. 
\end{itemize}

\par
In the above analysis, we found that the F444W image is more effective than the F115W or F200W images in detecting the moderate size stellar bar present in the galaxy CEERS-30155 at a redshift of $z\sim2.136$. However, ellipse fits of F444W images can only robustly detect bars whose sma is at least one PSF ($\sim0\farcs16$ corresponding to 1.4 kpc at $z\sim2$). Therefore, in general, F444W images will not be able to robustly detect stellar bars smaller than 1.5 kpc, which may be abundant at earlier epochs when galaxy disks were smaller. One approach to this challenge is to use a combination of shorter-wavelength images (e.g., F200W) and longer-wavelength images (e.g., F444W) to improve bar detection at $z>2$. The shorter-wavelength images have a higher spatial resolution, while the longer-wavelength images trace longer rest-frame wavelength light, which is a better tracer of low-mass, long-lived stellar populations in the stellar bar structure. We note that \citealt{Guo-etal-2024} applied this approach and found that the bar fraction derived from the combination of F200W and F444W images is higher than that obtained from either image individually (Figures 5 \& 6 in \citealt{Guo-etal-2024}).

\par
While \textit{JWST} F200W and F444W images can probe stellar bars at $z\sim2$, these images would encounter limitations at higher redshifts. Going to even earlier epochs ($z>4$), F200W images will begin to trace rest-frame UV light, which is highly obscured by dust. As redshift increases, stellar bars are likely to become smaller because galaxy disks become smaller, and the PSF of both F200W and F444W images will become increasingly inadequate for resolving these stellar bars. To overcome this limitation and improve stellar bar detection in higher redshift galaxies, the Giant Magellan Telescope (\textit{GMT}) would be a promising tool to use. The \textit{GMT} will be the largest ground-based infrared telescope with a diameter of 25.4 meters, significantly improving the PSF with the use of adaptive optics.

\bigskip

\acknowledgements K.P.J gratefully acknowledges support from the NSF REU grant AST-2244278 (PI: Jogee).

\bibliography{aspauthor}

\end{document}